\documentstyle[prb,aps,multicol,epsf]{revtex}

\begin{document}
\draft

\title{Single hole spectral function and spin-charge separation in the $t-J$ model}

\author{A. S. Mishchenko$^{1,2}$, N. V. Prokof'ev$^{3}$, and 
B. V. Svistunov $^{2,3}$}

\address{
$^1$Correlated Electron Research Center,
Tsukuba 305-0046, Japan  \\
$^2$ Russian Research Center ``Kurchatov Institute", 123182 Moscow, 
Russia  \\
$^3$ Department of Physics, University of Massachusetts, 
Amherst, MA 01003, USA \\
 }

\maketitle
\begin{abstract}
Worm algorithm Monte Carlo simulations of the hole Green 
function with subsequent spectral analysis were performed 
for $0.1 \le J/t \le 0.4$ on lattices with 
up to $L \times L = 32 \times 32$ 
sites at temperature as low as  $T=J/40$, and present, apparently, 
the hole spectral function in the thermodynamic limit. Spectral analysis 
reveals a $\delta$-function-sharp quasiparticle peak at the lower edge 
of the spectrum which is incompatible with the power-law singularity
and thus rules out the possibility of spin-charge separation in 
this parameter range. Spectral continuum features two peaks 
separated by a gap $\sim 4 \div 5~t$. 
\end{abstract}

\begin{multicols}{2}
\narrowtext

For almost four decades the problem of hole dynamics 
in magnetic systems has attracted constant interest with applications
ranging from properties of charge carriers in magnetic semiconductors and insulators
\cite{deGennes,Bulaevskii} to vacancies in solid $^3$He \cite{Andreev}.  
The research in this area exploded  with the discovery of high temperature
superconductors in cuprates, where superconductivity appears upon light doping of
AFM insulators.
Despite an enormous theoretical effort
over the years and quite a variety of treatments (for reviews, see, e.g.,
Ref.~\onlinecite{Manousakis91})
a complete solution of this 
inherently strong-coupling problem still does not exist,
especially in the most interesting
 region of $t > J$, where $J$ is the 
exchange coupling constant and  $t$ is the hopping matrix element in the 
$t-J$ Hamiltonian
\begin{equation}
H=-t\sum_{<ij>, s } c_{is}^{\dag}\: c_{js } + 
J\sum_{<ij>} \left( {\bf s}_i \cdot  {\bf s}_j -{1\over 4} n_i n_j \right) \;. 
\label{tJ}
\end{equation}
Here $c_{j\sigma}$ is projected (to avoid double occupancy) fermion 
annihilation operator,
$n_i= \sum_{s} c_{i s}^{\dag}\: c_{i s} \ne 2 $ is the occupation number, 
$ {\bf s}_i =
\sum_{ss'} c_{i s}^{\dag}\: {\bf \sigma}_{ss'} c_{i s'}$
is spin-$1/2$ operator, and $<ij>$ denote nearest neighbor sites of the 
2D square lattice. 

The central problem in the hole dynamics is whether or not its spin and
charge degrees of freedom separate. The standard  way to answer this question is
to study the 
spectral function $A_{\bf p} (\omega) = -\pi ^{-1} Im~ G_{\bf p} (\omega)$, where 
$G_{\bf p} (\omega)$ is the hole Green function. If there is an
elementary excitation associated with the hole, the spectral function is
supposed to feature a peak at the lower edge of the spectrum.  What is
crucial, however, is not
the presence of the peak itself, but its functional form \cite{example}.
Within the self-consistent Born approximation scheme (SCBA) \cite{Schmitt} one finds
finite overlap between the bare hole and low-energy quasiparticle states, 
which means that the peak is $\delta$-functional and 
the hole is described as coherently
propagating spin-polaron  in the nearly ordered antiferromagnetic (AFM) background
(with vanishing scattering at low temperature due to 
small density of spin waves) \cite{Schmitt}.
 
In contrast to that, various 
resonating-valence-bond (RVB) descriptions and Anderson's general arguments about 
breakdown of the  Fermi-liquid picture in the system with no-double-occupancy constraint
(see, e.g., Ref.~\onlinecite{Anderson}) strongly suggest that power-law singularity,
which is indicative of spin-change separation scenario, might be the case (there is even a
claim that the quasiparticle weight $Z$ should be rigorously zero \cite{Weng}).
To make the issue more confusing,
angle resolved photoemission spectroscopy experiments \cite{Wells,Ronning} show very broad
maximum in $A(\omega )$ which can be considered both as the  quasiparticle peak
with anomalously large broadening or as the evidence for composite
nature of quasiparticles \cite{Laughlin1}. 

The quasiparticle picture 
was supported by exact calculations on small 
clusters \cite{Manousakis91,Laughlin2,Leung}, but system sizes  
(up to $32$ sites) were too small to perform finite-size scaling. 
Variational calculations, Green function Monte Carlo and density matrix renormalization group 
studies were mostly concerned with the dispersion law $\epsilon_{\bf k}$ (lowest energy
in a given momentum sector).
Large scale simulations of the imaginary time Green function
$G_{\bf k} (\tau )$ were performed recently using a combination of the 
loop-cluster Monte Carlo method for the AFM state and hole evolution in the fixed 
space-time spin background \cite{Brunner}. This method works only for relatively large 
exchange $J>0.6~t$, since magnetic background is simulated {\it without} the hole
and polaron-type distortions have to be accounted for as quantum fluctuations {\it before}
the hole is introduced.
For $J/t < 0.6$ the error bars in $G_{\bf k} (\tau )$ are too large for reliable
spectral analysis (see below). 

To summarize, we still lack evidence    
that for small $J$ the quasiparticle weight remains finite in the thermodynamic limit 
and the lowest peak has nothing to do with the power-law singularity.  
We thus find it important to rule out the possibility that t-J model may explain
the data of Refs.~\onlinecite{Wells,Ronning} (as suggested by 
Ref.~\onlinecite{Weng}), so that extensions of the model such as $t'$ and $t''$ terms
\cite{Lee} or frustrating exchange couplings are proven necessary. 

Speaking classically, moving hole breaks AFM bonds and thus its energy
increases linearly with the travel distance \cite{Bulaevskii,Siggia} (this consideration, or the
string-potential picture, is most appropriate for the $t-J_z$ model). 
It is believed that the ground energy scaling $E_{{\bf k}_0} \sim J^{2/3}$ 
[where ${\bf k}_0=(\pi /2, \pi/2)$]
and excitation spectrum
are described by the string-potential picture 
\cite{Schmitt,Manousakis91,Manousakis92,Laughlin2,Brunner}, and transverse spin fluctuations
do not ``erase'' strings completely. In the limit of small $J$ the theory predicts
that several resonances in $A(\omega )$ have to be seen with the peak positions being
strictly related to the eigenvalue properties of Airy functions. 
Early exact diagonalization studies  on clusters $4\times 4$  
\cite{Manousakis91} attributed two peaks above the ground state to string resonances,
however later studies on larger clusters \cite{Leung,Laughlin2} were not able to detect
the second resonance, and the spectral function was showing strong size dependence.
What happens at small $J$ in the thermodynamic limit remains
an open question.    
 
In this letter we present results for $G_{\bf k} (\tau )$
and $A_{\bf k} (\omega)$ obtained from Monte Carlo simulations 
on systems with $16\times 16$ and $32 \times 32$ sites and at temperatures
as low as $T/J=0.025$ (for the largest system size) using 
continuous-time Worm algorithm \cite{ourJETP} in combination with the recently
developed spectral analysis which is capable of resolving infinitely sharp features
in $A(\omega)$ \cite{polaron}. 
The method itself is 
free from any systematic errors, and we were unable to detect
finite-size corrections in our data; thus, we believe, our results describe 
correctly the thermodynamic limit. In the parameter range studied 
$0.1 \le J/t \le 0.4$, the lowest peak in $A_{{\bf k}_0} (\omega)$
is a $\delta$-function within the resolution limit of order of $0.01 \div 0.03~t$,
which means that our quasiparticle is the spin-polaron. 
For the excitation spectrum we observe two well-separated peaks 
for all values of $J$. The ground state energy scaling does follow the
$J^{2/3}$ law predicted by the string-potential theory. Although we were 
unable to resolve individual string resonances, we believe that their combined effect
is seen as the first peak in the spectral continuum since its position also scales as
$J^{2/3}$. The high-energy peak is roughly at a constant distance $\sim 5t$  
from the ground state.

Worm algorithm is based on the idea that    
world-line configurations of spins and the hole
are updated through the space-time motion of the creation and annihilation 
operators. In Fig.~1 and Fig.~2 we show the typical configuration
of the Heisenberg
AFM with the hole, and Monte Carlo updates which apply to it. Physical 
configurations contributing to the hole Green function are those which have no
spin end points (denoted by filled circles).

Since the world-line representation      
is based on the expansion of the statistical evolution operator $e^{-H/T}$ 
in powers of $t$ and $J$ it suffers from the sign
problem which first appears in order $t^2 J^3 $ (see Fig.~3).
It is worth noting that if not for the sign problem,
spin-charge separation can be ruled out
by the analysis of world-line configurations. 
Let the hole be created by $c_{i\downarrow}$ 
operator. If spin charge separation does take place,
one should see, following the evolution of the system configuration
in imaginary time,
an extra spin density leaving the hole creation site and going into the bulk.
i.e., the world-line density far from the hole should increase.
At $T=0$ the AFM ground state is ordered
(as opposite to the spin liquid
state) and the minimal possible change in the world-line density 
is equivalent to having exactly  one extra world

%%%%%%%%%%%%%%%%%%%%%%%%%%%%%%%%%%%%%%%%%
\begin{figure}
\begin{center}
\epsfxsize=0.38\textwidth
\epsfbox{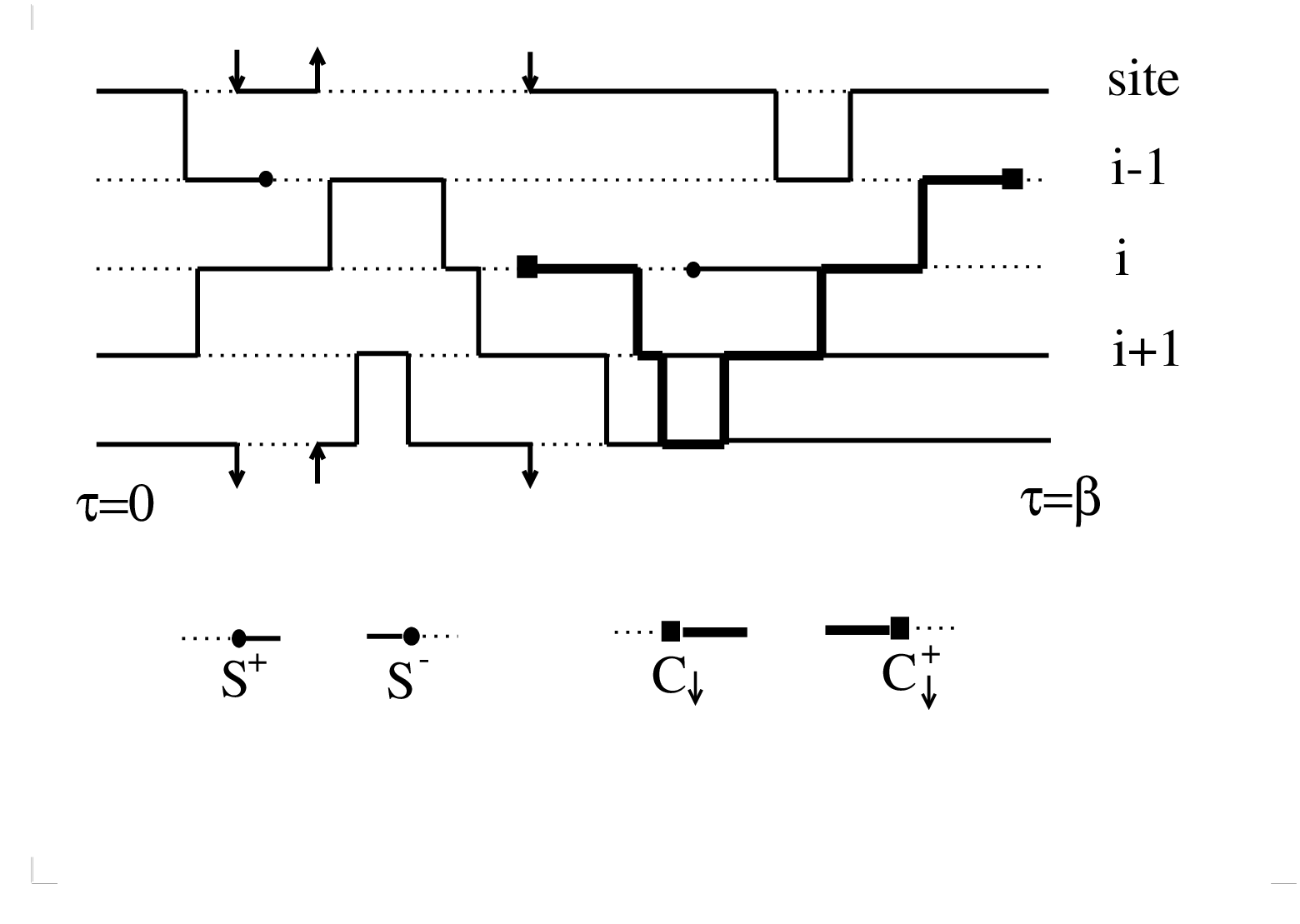}
\end{center}
\vspace*{-5.cm}
\caption{World-line configuration describing
quantum AFM with the hole; solid (dotted) lines correspond to
spin up (down) states and the bold line describes the hole.
Arrows indicate how 
periodic boundary conditions are used.  }
\label{fig:fig1}
\end{figure}
%670.00 670.00 scale
%0 rotate
%-0.05 0.35 translate
%%%%%%%%%%%%%%%%%%%%%%%%%%%%%%%%%%%%%%%%%

\vspace{0.cm}
%%%%%%%%%%%%%%%%%%%%%%%%%%%%%%%%%%%%%%%%%
\begin{figure}
\begin{center}
\epsfxsize=0.38\textwidth
\epsfbox{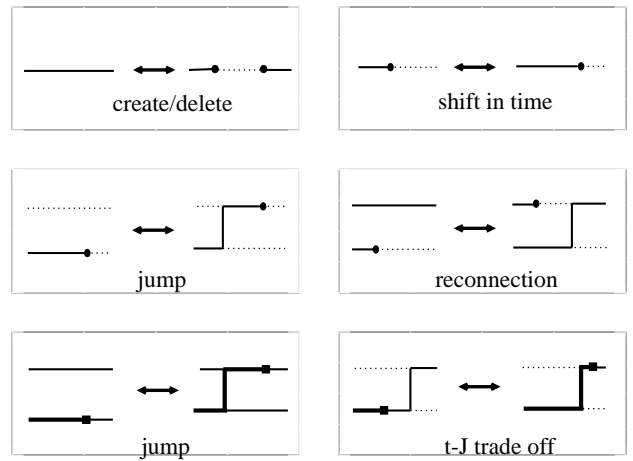}
\end{center}
\vspace*{-3.2cm}
\caption{Elementary Monte Carlo updates which form an ergodic
set. We show updates for $s^+$ and $c_{\uparrow}^{\dag}$ end points only
since procedures for $s^-$ and $c_s^{\dag}$, $c_s$ are identical up to a change
of notations (i.e., using proper incoming and outgoing lines).  
}
\label{fig:fig2}
\end{figure}
%670.00 670.00 scale
%0 rotate
%-0.05 0.3  translate
%%%%%%%%%%%%%%%%%%%%%%%%%%%%%%%%%%%%%%%%%

\vspace{-1.6cm}
%%%%%%%%%%%%%%%%%%%%%%%%%%%%%%%%%%%%%%%%%
\begin{figure}
\begin{center}
\epsfxsize=0.33\textwidth
\epsfbox{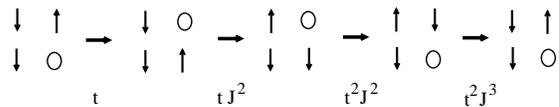}
\end{center}
\vspace*{-5.0cm}
\caption{The lowest order (in $t$ and $J$) identical transformation
leading to the sign problem: Corresponding graphical element changes the sign
of the configuration.
}
\label{fig:fig3}
\end{figure}
%%%%%%%%%%%%%%%%%%%%%%%%%%%%%%%%%%%%%%%%%

\noindent
 line, which can 
be traced out and interpreted as spin-$1$ magnon excitation.  
We may now construct an operator which has finite overlap 
with the quasiparticle excitation as a product 
$s^- c_{\downarrow}$, where $s^-$
is added to cancel the extra magnon in the bulk. However, up to
a single hopping transition the above composite operator is identical to 
$c_{\uparrow}$, and we conclude that holes are
good quasiparticles in contradiction with original assumption 
(probably rephrasing the proof of Ref.~\onlinecite{Schmitt}).   

However, in the presence of the sign problem the above consideration should 
be taken with extreme caution. It may turn out that
ordered world-line configurations compensate each other completely and 
single-configuration conclusions are misleading, as suggested in 
Ref.~ \onlinecite{Weng} where hole-related sign problem is called the 
``irreparable phase string effect'' and argued to cause
spin-charge separation. 

The sign problem implies that we may not
calculate $G(\tau )$ reliably over long time scales and have to restrict 
our simulation to $t \tau < 3 \div 4$ to suppress sign fluctuations by
larger statistics. Fortunately, on this time scale $G_{\bf k} (\tau)$ is already 
in its asymptotic regime and the data are sufficiently accurate to reveal the
ground state properties. Formally, calculations are done at finite 
$T$ but its value is more than an order of magnitude smaller than the 
energy of the lowest magnon state in a given system size. For each value of J
the calculation time was about 2 weeks on a PIII-600 workstation.

In Fig.~4 we show simulated $G_{{\bf k}_0} (\tau  ) $ for $J/t=0.4$ and the asymptotic law
$Z_{{\bf k}_0} e^{-E_{{\bf k}_0} \tau}$ with the quasiparticle weight
and ground state energy obtained from the weight and position of the $\delta$-peak
in $A_{{\bf k}_0} (\omega )$. 
Note, that for small values of $J$ the data have to be very accurate
to describe correctly how $G(\tau )$ approaches its asymptotic 
behavior $G_{{\bf k}_0} \to Z_{{\bf k}_0} e^{-E_{{\bf k}_0} \tau}$.
Error bars are shown but are smaller than the symbol size 
(the relative accuracy is better than 
$10^{-2}$ even for points with the largest
$\tau $ where the sign problem was the most severe).

The spectral analysis of $G_{{\bf k}_0} (\tau  ) $ was done using stochastic
optimization procedure developed earlier for the polaron problem.
$A(\omega )=N^{-1} \sum_{i=1}^N A_i(\omega )$ 
is obtained as an average over spectral densities which optimize
deviations between $G(\tau ) $ and 
$\int d\omega e^{-\omega \tau } A_i(\omega )$. The parameter space      
of $ A_i(\omega )$ is defined by the step-wise constant functions,
which, in particular, includes infinitely sharp peaks and is not
associated with any predefined set of frequencies  \cite{polaron}
(it is known that $\delta$-peaks can not be handled satisfactorily
by the maximum entropy method \cite{Brunner,Jarrell}). 

In Fig.~5 we show our results for $A_{{\bf k}_0} (\omega )$
calculated at points $J/t=0.4,~0.2,~0.1$.     
We clearly see a $\delta$-sharp peak at the lower edge of the spectrum.
The structure of this peak is incompatible
with the power law singularity since its width is smaller than
the lowest magnon excitation in our system [for $J/t =0.4$ the
quasiparticle peak width is only $0.01~t$ (!) while the natural scale
for the power law is set by $J$].
This is the central result 
of our paper which conclusively rules out spin-charge separation
scenario for the single hole dynamics in the $t-J$ model and confirms
finite quasiparticle weight in the thermodynamic limit. 
To verify that finite-size and finite-temperature corrections
are negligible we performed long time simulations 
in a $32 \times 32$ 
lattice at temperature $T=J/40$ for $J/t=0.2$, but within the error bars 
$G(\tau )$ was indistinguishable from the result obtained for 
$L=16$ and $T=J/20$. 

Unfortunately, the ill-defined problem of numeric analytic continuation
does not allow to study fine structures in the spectral
density, especially if they are ``screened'' 
by low- and high-frequency peaks. (The low-frequency 

\vspace*{-0.2cm}
%%%%%%%%%%%%%%%%%%%%%%%%%%%%%%%%%%%%%%%%%
\begin{figure}
\begin{center}
\epsfxsize=0.38\textwidth
\epsfbox{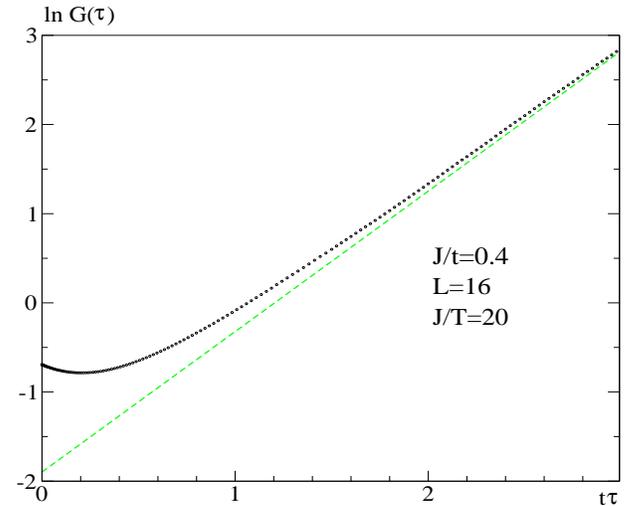}
\end{center}
\vspace*{-2.2cm}
\caption{ $G_{{\bf k}_0} (\tau  )$ (circles) and the asymptotic line
$Z_{{\bf k}_0} e^{-E_{{\bf k}_0} \tau}$ (dashed) for J/t=0.4.
}
\label{fig:fig4}
\end{figure}
%690.00 620.00 scale
%0 rotate
%-0.05 0.260 translate
%%%%%%%%%%%%%%%%%%%%%%%%%%%%%%%%%%%%%%%%%

\vspace*{0.3cm}
%%%%%%%%%%%%%%%%%%%%%%%%%%%%%%%%%%%%%%%%%
\begin{figure}
\begin{center}
\epsfxsize=0.382\textwidth
\epsfbox{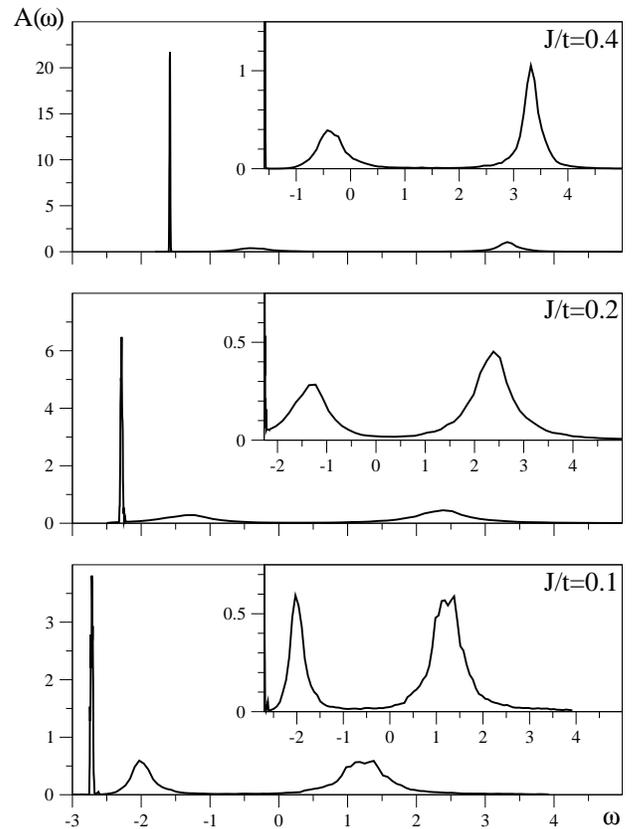}
\end{center}
\vspace*{1.4cm}
\caption{ Spectral functions for $J/t=0.4,~0.2$ and $0.1$. Frequency
is measured in units of $t$ and the integral $\int d\omega A(\omega )$ is normalized 
to unity. These spectra were obtained for the $16 \times 16$ lattice  at $T=J/20$.}
\label{fig:fig5}
\end{figure}
%  760.00 760.00 scale
%  0.03 -0.18 translate
%%%%%%%%%%%%%%%%%%%%%%%%%%%%%%%%%%%%%%%%%

\noindent
peak is fixed by the asymptotic
long-time 
behavior of $G(\tau )$, while the high-frequency peak is
fixed by the short-time decay of $G(\tau )$ where the data are 
extremely accurate.) 
Our tests show that multiple peaks in the middle
can not be resolved by spectral analysis even when we use 
analytically exact $G(\tau )$ data. It means that the absence of multiple
string resonances above the ground state in our results for $A(\omega )$
{\it may not } be considered as a proof that string potential picture fails
in quantum case. We would rather consider the second peak as a ``course grain''
description of spectral density at intermediate energies. However, if string
excitations do exist, their combined effect should be seen as the $J^{2/3}$ scaling
law for the peak position. In Fig.~6 we plot peak positions as functions of
$(J/t)^{2/3}$ for $0.1 \le J/t \le 1.2$ with error bars obtained as 
peak half-widths. We conclude that for the second peak 
the scaling law is obeyed within the  error bars. 
The high-energy peak stays roughly at a constant distance from the ground
state, and clearly the physics behind it is different. 

\vspace{0.3cm}
%%%%%%%%%%%%%%%%%%%%%%%%%%%%%%%%%%%%%%%%%
\begin{figure}
\begin{center}
\epsfxsize=0.375\textwidth
\epsfbox{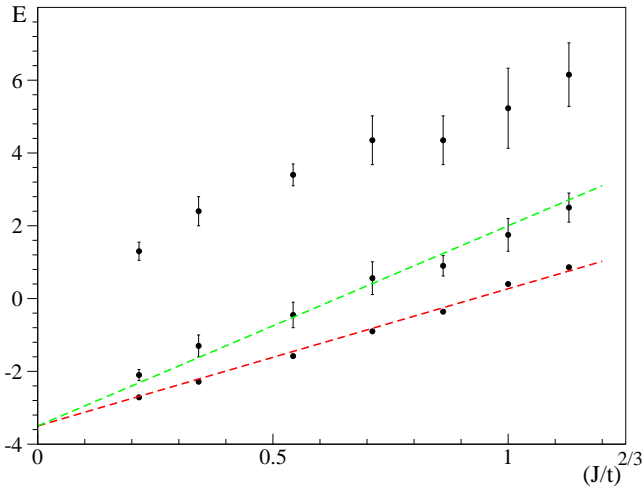}
\end{center}
\vspace*{-2.9cm}
\caption{Peak positions as functions of $(J/t)^{2/3}$. Data points for
$J>0.4$ were taken from Ref.~[15] (for $J=0.4 ~t$ the second peak was not
resolved in  Ref.~[15] because of large error bars in $G(\tau )$).
The two lines are fits 
$y(x)=a+b (J/t)^{2/3}$ with $a=-3.5~t$, $b=3.77~t$ for the ground state, and
$b=5.5~t$ for the first peak in continuum.
}
\label{fig:fig6}
\end{figure}
%630.00 630.00 scale
%-0.0 0.35 translate
%%%%%%%%%%%%%%%%%%%%%%%%%%%%%%%%%%%%%%%%%

We thank O. Ruebenacker, P. Stamp, and V. Kashurnikov for 
valuable discussions. 
This work was supported by the National Science Foundation under Grant
DMR-0071767.

\end{multicols}
        
\end{document}